\documentclass[10pt,conference]{IEEEtran}

\usepackage{amsfonts}
\usepackage{graphicx}
\usepackage{textcomp}
\usepackage{xcolor}
\usepackage[linesnumbered,lined,ruled,commentsnumbered]{algorithm2e}
\usepackage{xspace}
\usepackage{color}
\usepackage{multirow}
\usepackage{lscape} 
\usepackage{subcaption}
\usepackage{enumitem}
\usepackage{listings, xcolor, soul} 
\usepackage{colortbl}
\usepackage[switch]{lineno}
\usepackage[T1]{fontenc}
\usepackage{textcomp}
\usepackage{adjustbox}

\usepackage{amsmath}
\usepackage{amssymb}
\usepackage{amsthm}

\usepackage{booktabs}
\usepackage{hyperref}
\usepackage{tcolorbox}
\usepackage{bbding}

\setlist[itemize]{noitemsep, topsep=4pt}
\definecolor{lightgray}{HTML}{f6f6f6}
\definecolor{darkgray}{rgb}{.4,.4,.4}
\definecolor{darkblue}{HTML}{800080}
\definecolor{brickred}{HTML}{b04f4f}
\definecolor{purple}{rgb}{0.65, 0.12, 0.82}
\definecolor{diffadd}{HTML}{288f26}
\definecolor{diffrmbg}{HTML}{ffebe9}
\definecolor{diffaddbg}{HTML}{e6ffeb}
\definecolor{diffremove}{HTML}{de4f54}
\definecolor{carrotorange}{rgb}{0.8, 0.33, 0.0}
\definecolor{highlight}{HTML}{fefbc2}
\definecolor{bluegray}{HTML}{3182bd}
\definecolor{delim}{RGB}{20,105,176}
\definecolor{numb}{RGB}{106, 109, 32}
\definecolor{string}{rgb}{0.64,0.08,0.08}
\definecolor{geminipurple}{HTML}{D0BCFF}
\colorlet{punct}{red!60!black}
\definecolor{delim}{RGB}{20,105,176}
\colorlet{numb}{magenta!60!black}

\lstdefinelanguage{Python}{
	keywords={typeof, new, true, false, catch, function, return, null, catch, switch, var, const, let, extends, if, in, while, do, else, case, break, async, await, of, from, import, class, def},
	keywordstyle=\color{darkblue}\bfseries,
	ndkeywords={class, export, boolean, throw, implements, import, this, setTimeout, self, __init__},
	ndkeywordstyle=\color{brickred}\bfseries,
	identifierstyle=\color{black},
	sensitive=false,
	comment=[l]{//},
	morecomment=[f][\color{diffadd}\bfseries]{+\ },
	morecomment=[s]{/*}{*/},
	morecomment=[f][\color{diffremove}\bfseries]{- },
	commentstyle=\color{violet}\ttfamily,
	stringstyle=\color{carrotorange}\ttfamily,
	morestring=[b]',
	morestring=[b]"
}

\lstdefinelanguage{json}{
	basicstyle=\normalfont\ttfamily\small,
	numbers=left,
	numberstyle=\scriptsize,
	stepnumber=1,
	numbersep=8pt,
	showstringspaces=false,
	breaklines=true,
	frame=single,
	backgroundcolor=\color{gray!5},
	literate=
	*{0}{{{\color{numb}0}}}{1}
	{1}{{{\color{numb}1}}}{1}
	{2}{{{\color{numb}2}}}{1}
	{3}{{{\color{numb}3}}}{1}
	{4}{{{\color{numb}4}}}{1}
	{5}{{{\color{numb}5}}}{1}
	{6}{{{\color{numb}6}}}{1}
	{7}{{{\color{numb}7}}}{1}
	{8}{{{\color{numb}8}}}{1}
	{9}{{{\color{numb}9}}}{1}
	{:}{{{\color{punct}{:}}}}{1}
	{,}{{{\color{punct}{,}}}}{1}
	{\{}{{{\color{delim}{\{}}}}{1}
	{\}}{{{\color{delim}{\}}}}}{1}
	{[}{{{\color{delim}{[}}}}{1}
	{]}{{{\color{delim}{]}}}}{1},
}

\let\origthelstnumber\thelstnumber
\makeatletter
\newcommand*\Suppressnumber{%
	\lst@AddToHook{OnNewLine}{%
		\let\thelstnumber\relax%
		\advance\c@lstnumber-\@ne\relax%
	}%
}

\definecolor{Gray}{gray}{0.6}
\definecolor{LightGray}{gray}{0.9}
\definecolor{bananayellow}{rgb}{1.0, 0.88, 0.5}
\definecolor{myred}{rgb}{1.0,0.44,0.37}

\newcolumntype{a}{>{\columncolor{LightGray}}c}
\newcolumntype{b}{>{\columncolor{LightGray}}r}

\newcommand*\Reactivatenumber[1]{%
	\setcounter{lstnumber}{\numexpr#1-1\relax}
	\lst@AddToHook{OnNewLine}{%
		\let\thelstnumber\origthelstnumber%
	}%
}

\makeatother

\SetCommentSty{mycommfont}

\theoremstyle{definition}

\newcommand{\header}[1]{\par\noindent\textbf{#1.}}

\def\BibTeX{{\rm B\kern-.05em{\sc i\kern-.025em b}\kern-.08em
		T\kern-.1667em\lower.7ex\hbox{E}\kern-.125emX}}

\newboolean{showcomments}
\setboolean{showcomments}{true}
\ifthenelse{\boolean{showcomments}}
{
	\definecolor{myyellow}{RGB}{255, 228, 26}
	\definecolor{myblue}{RGB}{50, 50, 220}
	\newcommand{\nb}[2]{
		{\sf
			\fcolorbox{myyellow}{yellow}{\scriptsize\textbf{#1}}%
			$\blacktriangleright$%
			{\color{myblue}\fontsize{7pt}{8pt}\selectfont\textbf{#2}}%
		}%
	}
}
{
	\newcommand{\nb}[2]{}
}

\newcommand{\panta}{\textsc{Panta}\xspace}
\newcommand{\symprompt}{\textsc{SymPrompt}\xspace}
\newcommand{\toolname}{\textsc{Scate}\xspace}
\newcommand{\beacon}{\textsc{Scate MCP}\xspace}

\newcommand{\claudecode}{\textsc{Claude Code}\xspace}
\newcommand{\codex}{\textsc{Codex CLI}\xspace}
\newcommand{\geminicli}{\textsc{Gemini-cli}\xspace}
\newcommand{\qwencode}{\textsc{Qwen Code}\xspace}

\newcommand{\code}[1]{{\small\ttfamily\texttt{#1}}}    

\makeatletter
\DeclareRobustCommand{\change}{%
	\@bsphack
	\leavevmode
	\color{blue}
	\@esphack
}
\DeclareRobustCommand{\stopchange}{%
	\@bsphack
	\normalcolor
	\@esphack
}
\makeatother

\begin{document}


\title{
\toolname: Learning to Supervise Coding Agents for Cost-Effective Test Generation
}
        
\author{
	\IEEEauthorblockN{Sijia Gu}
	\IEEEauthorblockA{\textit{University of British Columbia}\\
		Vancouver, Canada \\
		sijiagu@ece.ubc.ca}
	\and
	\IEEEauthorblockN{Noor Nashid}
	\IEEEauthorblockA{\textit{University of British Columbia}\\
		Vancouver, Canada \\
		nashid@ece.ubc.ca}
	\and
	\IEEEauthorblockN{Ali Mesbah}
	\IEEEauthorblockA{\textit{University of British Columbia}\\
		Vancouver, Canada \\
		amesbah@ece.ubc.ca}
}
\maketitle

\thispagestyle{plain}
\pagestyle{plain}

\begin{abstract}
While autonomous coding agents have significantly advanced automated test generation, they remain fundamentally limited by \textit{lazy generation}, a phenomenon where agents prematurely terminate tasks and systematically avoid complex programmatic logic, resulting in inadequate code coverage. Currently, mitigating this premature termination requires continuous human-in-the-loop supervision. This heavy reliance on human intuition creates a bottleneck that negates the efficiency gains of automated generation. We propose \toolname, a framework for adaptive, automated supervision of coding agents that replaces human intervention during test generation. 
By formulating supervision as a contextual bandit problem, \toolname learns to select the most promising testing actions based on the current coverage and class testability metrics, maximizing coverage gains while minimizing wasted generation effort.
Our empirical evaluation demonstrates that 
\toolname integrates seamlessly with different coding agents. When applied to \geminicli, it achieves 32.3\% higher line coverage and 30.9\% higher branch coverage than the agent-only baseline. A comparison with \claudecode confirms the framework dynamically adapts its policy to optimize each agent's unique strengths.
\toolname also consistently outperforms state-of-the-art non-agentic approaches across all metrics. 
\end{abstract}

\section{Introduction}\label{intro}
The rapid evolution of Large Language Models (LLMs) has led to their widespread adoption in automated test generation tasks \cite{wang2024hits, symprompt, panta, altmayer2025coverup, utgen, pan2025aster}. Recent advances have further enabled the emergence of \emph{coding agents} such as  \codex \cite{openai-codex}, \claudecode ~\cite{claude-code}, \geminicli~\cite{gemini-cli}, and \qwencode ~\cite{qwen-code}. Coding agents are terminal-based, LLM-driven systems capable of iteratively interacting with software projects through actions such as searching, navigating, editing, building, and testing. Unlike LLM prompting approaches \cite{wang2024hits, symprompt, panta, altmayer2025coverup, utgen, pan2025aster}, which rely on heuristic, rule-based generation strategies, coding agents operate autonomously over multiple iterations, adapting their behaviour based on intermediate feedback from compilers, test execution outcomes, and the surrounding source code context.

Despite this autonomy, state-of-the-art coding agents suffer from critical limitations when applied to comprehensive test generation. A primary challenge is the phenomenon of \emph{lazy generation}~\cite{ivanov2026can}. Even when explicitly instructed to maximize code coverage, agents naturally gravitate toward simple methods and straight-line execution paths, systematically avoiding complex logic that requires deep program analysis. To illustrate this, \autoref{fig:lazy-generation-example} demonstrates a real-world interaction with the \geminicli agent tasked with generating tests for Apache Commons CLI's \code{DefaultParser} class from Defects4J~\cite{just2014defects4j}. Despite the agent's confident report that tests are ``generated and verified", it prematurely terminates its execution, leaving branch coverage stalled at a mere 31\%.

\begin{figure}[htbp]
	\centering
    \scriptsize
	\begin{tcolorbox}[colback=white, coltext=black, fontupper=\ttfamily\scriptsize, title=Gemini CLI Output, sharp corners, top=2pt, bottom=2pt, left=4pt, right=4pt]
		> Generate tests for class DefaultParser to maximize its code coverage.\\
		\\
		{}[Agent Processing...]\\
		\\
		\scalebox{0.7}{\FourStar} Tests for \textbf{\textit{DefaultParser}} have been generated and verified [...] The tests successfully execute and have been integrated into the project's build system.\\
		\\
		> What's the current code coverage?\\
		\scalebox{0.7}{\FourStar} The current code coverage for the \textbf{\textit{DefaultParser}} class is approximately \textbf{42\% Line Coverage}, with \textbf{31\% Branch Coverage}. \\
		\\
		This indicates that [...] further testing is needed to improve coverage of edge cases and complex parsing logic.
	\end{tcolorbox}
	
	\caption{An example of test generation by Gemini CLI}
	\label{fig:lazy-generation-example}
\end{figure}

To address premature termination, developers often remain in the loop to steer coding agents, deciding whether to retry, provide targeted guidance, or halt the process. These decisions rely heavily on human intuition and can either leave critical branches untested or waste computational resources on ineffective iterations. Consequently, monitoring agent progress and providing corrective feedback becomes a bottleneck that undermines the efficiency gains of automated test generation.

In this paper, we introduce \toolname (\textbf{S}upervisor-\textbf{C}opiloted \textbf{A}gentic \textbf{T}est Generation \textbf{E}ngine), an adaptive supervisor designed to replace human-in-the-loop supervision. 
Unlike traditional automated testing frameworks that micro-manage generation at the granular level of individual methods or execution paths~\cite{symprompt, wang2024hits, panta}, \toolname employs a macro-level, class-centric strategy. 
By framing supervision as a contextual bandit problem~\cite{li2010contextual}, a sequential decision-making framework that learns to select actions based on the current context and observed rewards, \toolname dynamically combines runtime coverage feedback with structural metrics to provide continuous, data-driven guidance that directs coding agents toward comprehensive unit test generation.

In this paper, we make the following contributions:

\begin{itemize}

\item \toolname, to the best of our knowledge, the first framework to provide automated, adaptive supervision for coding agents in test generation. \toolname is agent-agnostic, readily integrates with diverse coding agents.

\item A novel formulation of agent supervision as a contextual bandit problem. By learning from static class testability and real-time code coverage metrics, the supervisor dynamically adapts its policy to maximize code coverage while minimizing computational costs.

\item An empirical evaluation on a dataset constructed from Defects4J open-source projects~\cite{just2014defects4j}. \toolname improves line and branch coverage over unsupervised coding agent baselines by 32.3\% and 30.9\%, respectively, with \geminicli, and by 6.0\% and 5.9\%, respectively, with \claudecode, demonstrating its effectiveness across diverse coding agents. It also consistently outperforms state-of-the-art non-agentic approaches across all evaluation metrics.

\end{itemize}

\section{Approach}
\begin{figure}
	\includegraphics[width=0.48\textwidth]{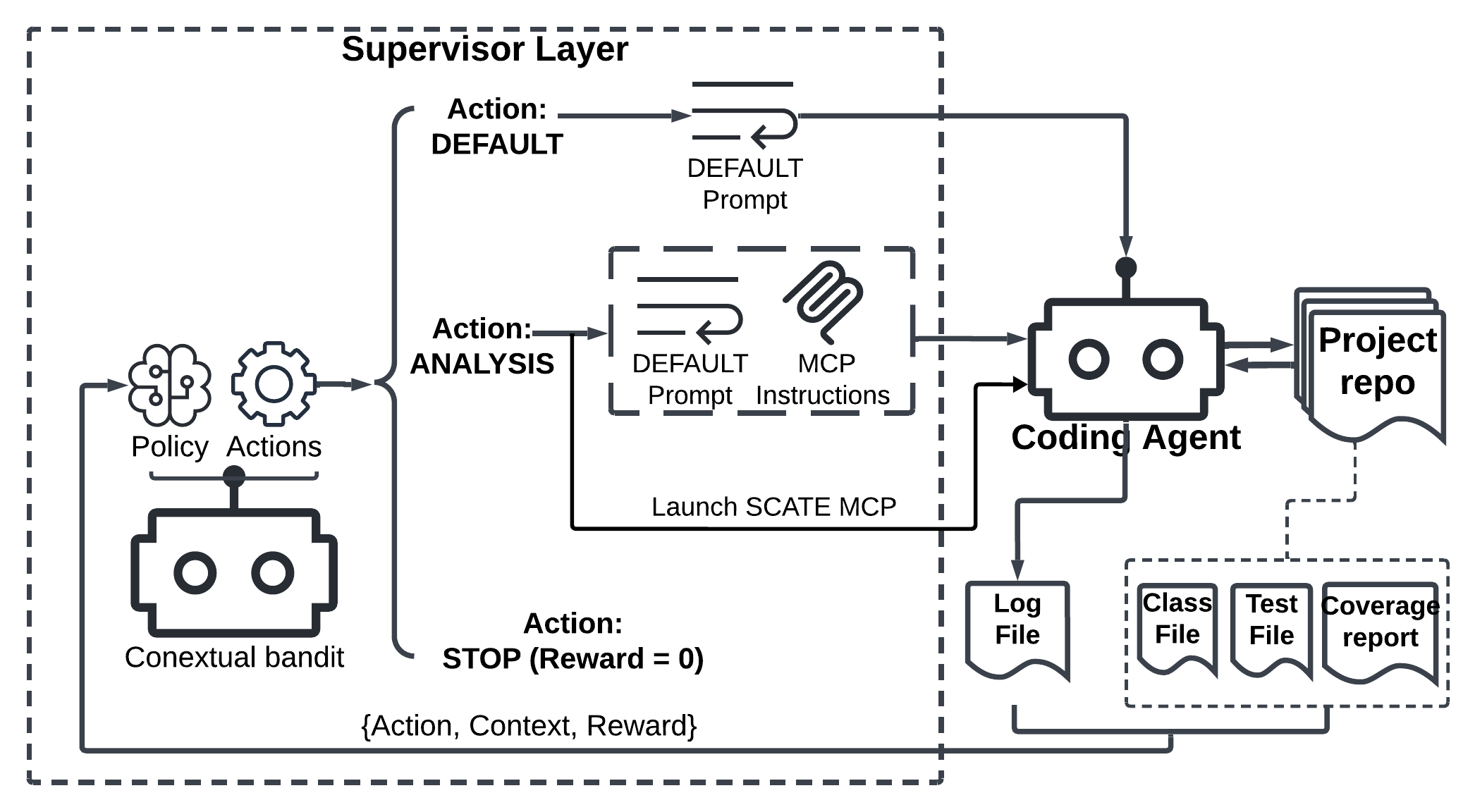}
	\caption{Overview of \toolname.}
	\label{overview}
\end{figure}

To fully automate unit test generation, \toolname employs a supervisor-orchestrated agentic workflow. The supervisor first assesses the target class using structural testability metrics. It then combines these static characteristics with dynamic feedback from the evolving test suite, including achieved coverage and uncovered complexity, to assess the current testing context and select the action most likely to improve test generation outcomes. At each timestep, the supervisor adapts its policy to balance coverage gains against computational cost, deciding whether to continue standard generation, invoke deeper program analysis for complex branches, or terminate the process to avoid resource exhaustion. \autoref{overview} provides an overview of the workflow. The following subsections describe the supervisor formulation, program analysis mechanisms, and supervisor-agent interaction.

\subsection{Contextual Bandit Formulation}

We formulate the \toolname supervisor as a contextual bandit problem~\cite{li2010contextual}. As a mature branch of reinforcement learning (RL)~\cite{sutton1998reinforcement}, the contextual bandit framework is highly effective at selecting optimal actions based on rich state observations~\cite{li2022reinforcement}, aligning directly with the context-aware design of \toolname. Unlike full RL formulations, which require modelling complex, sequential state transitions over thousands of training episodes, a contextual bandit efficiently learns a direct policy by optimizing for immediate, single-step payoffs. 
To formalize this approach, we first define the contextual features that capture the static and dynamic state of the target class, followed by the supervisor’s action space. We then present the reward function used to balance coverage gains against computational cost and conclude with the learning algorithm that derives the decision-making policy.

\subsubsection{Contextual Features}\label{approach:features}

Generating comprehensive unit tests requires LLM-based coding agents to understand source code, including complex control flows and extensive external dependencies, within a limited context window~\cite{panta, symprompt}.
The inherent complexity of a target class directly influences both achievable code coverage and the effort required to improve it. More complex classes are typically harder to test, achieve lower coverage, and require substantially greater agent effort.


We leverage class testability metrics to quantify the computational and reasoning demands of unit test generation~\cite{chidamber1994metrics, bruntink2006empirical, garousi2019survey, galindo2025increasing}. Although originally proposed to estimate human testing effort~\cite{toure2018predicting}, these metrics can serve as proxies for the challenges faced by coding agents. For example, high internal complexity increases the reasoning burden on the LLM, while high external coupling complicates environment setup and dependency mocking.

To construct a compact and informative state representation, we draw on the testability metrics identified by Bruntink et al.~\cite{bruntink2006empirical} and select a small set of aggregate, non-redundant features. Specifically, our state representation comprises three class-level metrics that characterize overall size, internal complexity, and external coupling:

\begin{itemize}
	\item \textbf{Lines of Code (LOC):} It acts as a baseline measure of class size, serving as a proxy for the foundational context window required by the LLM to parse the target class.
	\item \textbf{Weighted Methods per Class (WMC):} This metric represents the sum of the complexities of the methods within a given class, where each method is weighted by its cyclomatic complexity~\cite{McCabeCyclomatic}. It serves as a strong indicator of internal structural difficulty and the risk of agent reasoning failures.
	\item \textbf{Response for a Class (RFC):} Defined as the number of methods within the targeted class and the number of methods from other classes it invokes. This metric serves as a proxy for external dependencies and the complexity of the mocking required by the agent.
\end{itemize}

These metrics are computed using the control flow graph (CFG) of the target class. 
While static metrics characterize the inherent difficulty of the target class, the supervisor must also track generation progress. We therefore incorporate real-time line coverage ($Cov^{L}$), branch coverage ($Cov^{B}$), and \code{missed\_complexity}, defined as the number of uncovered method entries and decision points~\cite{jacoco}. Unlike relative coverage percentages, which can obscure the true amount of remaining work, \code{missed\_complexity} provides a direct measure of the unresolved structural testing opportunity. Thus, our resulting contextual feature vector comprises three static metrics (LOC, WMC, and RFC) and three dynamic metrics (line coverage, branch coverage, and \code{missed\_complexity}).

In practice, software structural metrics typically follow a power-law distribution. To ensure numerical stability and prevent classes with extreme complexity from disproportionately skewing the downstream learning process, we apply logarithmic scaling~\cite{normalization} to compress and normalize unbounded metrics as $\tilde{\text{\code{metric}}} = \ln(1 + \text{\code{metric}})/10$. At any discrete time step $t$, the complete contextual state observed by the supervisor is formally defined as the vector:

\begin{equation}\label{eq:context-vector}
\small
\mathbf{x}_t = \big[ 1, \tilde{L}, \tilde{W}, \tilde{R}, Cov^{L}, Cov^{B}, \tilde{M} \big]^\top
\end{equation}

Where:
\begin{itemize}
	\item $1$ acts as an intercept, ensuring baseline stability for the downstream decision policy.
	\item $\tilde{L}, \tilde{W}, \tilde{R}$ and $\tilde{M}$ represent the normalized LOC, WMC, RFC and \code{missed\_complexity} metrics, respectively.
	\item $Cov^L, Cov^B \in [0,1]$ represent the current line and branch coverage.
\end{itemize}

\subsubsection{Action Space}\label{approach:action}

Following the supervision challenge outlined in Section~\ref{intro}, we model agent guidance as a routing problem with three possible actions: \textit{Default}, \textit{Analysis}, and \textit{Stop}.
Constraining the action space to exactly these three macro-level choices is a deliberate mathematical and architectural design decision. A broader action space (e.g., forcing the supervisor to micro-manage individual analysis tools or prompt variations) would drastically expand the exploration space, slowing the contextual bandit's convergence and hindering its ability to efficiently discover coverage-maximizing strategies. Instead, these three arms encapsulate the entire intervention spectrum (baseline generation, targeted assistance, and zero-cost termination), enabling the supervisor to rapidly learn a cost-effective policy that maximizes coverage. We define the discrete action space as $\mathcal{A} = \{a_{\text{def}}, a_{\text{ana}}, a_{\text{stop}}\}$, representing the three available routing decisions:

\begin{enumerate}
    \item \textbf{Default ($a_{\text{def}}$):} A direct, zero-shot generation strategy. Relying solely on predefined project context.
    
    \item \textbf{Analysis ($a_{\text{ana}}$):} A tool-augmented generation strategy. The agent dynamically queries the \beacon tool, our custom program analysis tool deployed as a Model Context Protocol (MCP) ~\cite{mcp-spec} server, to navigate uncovered paths, resolve complex logical constraints, and identify mocking targets for complex code.
    
    \item \textbf{Stop ($a_{\text{stop}}$):} A termination signal to halt execution and save computational resources.
\end{enumerate}

Thus, while the Default action provides a rapid baseline for easily covered structures, the supervisor learns to dynamically allocate the tool-augmented Analysis action to complex classes where zero-shot generation reaches a plateau in capability. Ultimately, it chooses Stop to prevent resource exhaustion when further generation attempts yield diminishing returns. 

\subsubsection{Reward Function}\label{approach:reward}

Over a sequence of discrete timesteps $t = 1, 2, \dots, T$, where $T$ represents the total decision-making steps made by the supervisor, the system iteratively refines its policy. At each step $t$, after the supervisor evaluates the current contextual state vector $\mathbf{x}_t$ and selects an action $a_t \in \mathcal{A}$, the coding agent executes the corresponding generation strategy. The environment then returns an action-dependent reward signal $r_t$, providing critical feedback to update the supervisor's underlying policy. To achieve our objective of maximizing overall code coverage while minimizing cost (i.e., token usage), we design a reward function that formalizes this trade-off for every discrete action. 

Consistent with optimal stopping theory~\cite{ferguson2006optimal}, we define the reward for the \textit{Stop} action as zero ($r_t = 0$ when $a_t = a_{\text{stop}}$), reflecting a terminal state with neither additional cost nor benefit.
Conversely, the reward signal for active decisions (Default and Analysis) is designed to capture both the computational cost and the coverage gain incurred at step $t$. We calculate the relative coverage gain $g_t$, which prevents early-round bias by measuring the coverage improvement strictly relative to the remaining uncovered code. Let $Cov_{t-1}^{L}, Cov_{t-1}^{B} \in [0, 1]$ denote the initial line and branch coverage at the start of step $t$, and $Cov_{t}^{L}, Cov_{t}^{B} \in [0, 1]$ denote the coverage after the action is executed. The relative gain is defined as:
\begin{equation}\label{eq:relative-cov}
\small
g_t = \frac{1}{2} \left( \frac{Cov_t^L - Cov_{t-1}^L}{\max(1 - Cov_{t-1}^L, \epsilon)} + \frac{Cov_t^B - Cov_{t-1}^B}{\max(1 - Cov_{t-1}^B, \epsilon)} \right)
\end{equation}

where $\epsilon = 10^{-5}$ is a small constant ensuring numerical stability when coverage nears $1.0$. This formulation correctly values late-stage, complex test generation higher than raw absolute gains. For example, an absolute coverage increase of $0.05$ (from $0.90$ to $0.95$) yields a relative gain of $0.50$, representing the resolution of half the remaining uncovered code. 
By assigning greater rewards to late-stage coverage improvements, the formulation incentivizes the supervisor to pursue additional gains, such as transitioning from \textit{Default} generation to \textit{Analysis} when coverage growth begins to plateau.

As discussed in Section~\ref{approach:features}, achieving a given coverage level is substantially more difficult for complex classes than for simpler ones. To account for this disparity, we scale the relative coverage gain $g_t$ using a complexity multiplier, $m_t = 1 + \ln(1 + c_t)$, where $c_t = \max(0, Comp_t - Comp_{t-1})$ denotes the increase in covered complexity during step $t$, corresponding to the reduction in \code{missed\_complexity}. This formulation rewards progress on structurally complex classes while the logarithmic transformation preserves stable bandit updates.

To account for cost, each generation action incurs an API token cost, $Cost_t$, defined as the total monetary expense (in USD) associated with all input, output, and cached tokens extracted from the coding agent's log files. By default, $\omega = 1$, balancing complexity-scaled coverage gains and cost.

Finally, to discourage actions that do not improve coverage ($g_t \le 0$), we apply a fixed failure penalty ($\kappa = 0.5$). This value was empirically chosen to balance discouraging ineffective actions with maintaining exploration. The resulting action-dependent reward $r_t$ is defined as:

\begin{equation}\label{eq:reward}
\small
r_t = 
\begin{cases} 
0 & \text{if } a_t = a_{\text{stop}} \\
g_t \cdot m_t - \omega \cdot Cost_t & \text{if } a_t \neq a_{\text{stop}} \text{ and } g_t > 0 \\
-  \omega \cdot Cost_t - \kappa & \text{if } a_t \neq a_{\text{stop}} \text{ and } g_t \le 0 
\end{cases}
\end{equation}

\subsubsection{Contextual Bandit Algorithm}\label{approach:algorithm}

With the contextual state vector $\mathbf{x}_t$, action space $\mathcal{A}$, and reward function $r_t$ established, we operationalize the supervisor's decision-making policy using the \textit{LinUCB} algorithm~\cite{li2010contextual}. LinUCB is a highly efficient contextual bandit algorithm designed to dynamically balance the exploration of novel strategies with the exploitation of known successes. The core components of LinUCB are detailed in Algorithm \ref{alg:linucb_mechanism}.

The central optimization objective over the global horizon $T$ is to maximize the expected cumulative reward, denoted formally as $\mathbb{E} \big[ \sum_{t=1}^T r_t \big]$. To achieve this, the supervisor must consistently approximate the optimal action $a_t^*$ that yields the maximum expected reward at each discrete trial $t$. LinUCB facilitates this optimization by operating under the assumption~\cite{li2010contextual} that the expected reward for any action $a$ is a linear function of the context $\mathbf{x}_t$, parameterized by an unknown weight vector $\boldsymbol{\theta}^*_a$: $\mathbb{E}[r_{t}|a, \mathbf{x}_t] = \mathbf{x}_t^\top \boldsymbol{\theta}^*_a$. To learn these hidden parameters, the algorithm maintains a covariance matrix $\mathbf{A}_a \in \mathbb{R}^{d \times d}$ (where $d=7$ as defined in Section \ref{approach:features}) and a cumulative reward vector $\mathbf{b}_a \in \mathbb{R}^d$ for each action $a \in \mathcal{A}$. To ensure continuity across all timesteps, these structures ($\mathbf{A}$ and $\mathbf{b}$) are either loaded from persistent storage or, if no prior state exists, initialized as an identity matrix ($\mathbf{I}_d$) and a zero vector ($\mathbf{0}_d$) to provide an unbiased baseline (Lines \ref{line:initial-start}--\ref{line:initial-end}). During this iterative refinement, these matrices act as sufficient statistics: $\mathbf{A}_a$ records the outer products of past contexts ($\mathbf{x}_t \mathbf{x}_t^\top$) at Line \ref{line:update_A}, while $\mathbf{b}_a$ aggregates the context-weighted rewards ($r_t \mathbf{x}_t$) at Line \ref{line:update_b}. This state is saved to path $\mathcal{P}$ after each update to ensure persistence.

To govern \textit{action selection} at timestep $t$, the supervisor evaluates the current context $\mathbf{x}_t$ and computes the Upper Confidence Bound~\cite{sutton1998reinforcement} (UCB) score $p_{t,a}$ for each available strategy. Because the reward for the Stop action is a deterministic baseline ($r_t = 0$), its expected score carries no uncertainty and is strictly evaluated as $0.0$ (Line \ref{line:stop}). For the active generation strategies, the supervisor first derives the current empirical weight vector by solving the ridge regression estimate $\hat{\boldsymbol{\theta}}_a = \mathbf{A}_a^{-1}\mathbf{b}_a$ (Line \ref{line:theta}). It then calculates $p_{t,a}$ by combining this exploitative estimate ($\mathbf{x}_t^\top \hat{\boldsymbol{\theta}}_a$) with a mathematical exploration term ($\alpha \sqrt{\mathbf{x}_t^\top \mathbf{A}_a^{-1} \mathbf{x}_t}$), which acts as an uncertainty bonus scaled by exploration hyperparameter $\alpha$ (Line \ref{line:ucb_calc}). As the supervisor gathers more observations, the continuous refinement of $\mathbf{A}_a$ and $\mathbf{b}_a$ drives the empirical estimate $\hat{\boldsymbol{\theta}}_a$ closer to the true optimal parameters $\boldsymbol{\theta}^*_a$. Concurrently, the uncertainty bound geometrically shrinks, naturally shifting the supervisor's policy from broad exploration to confident exploitation. The final action $a_t$ with the highest overall UCB score is then selected (Line \ref{line:argmax}):

\begin{equation}\label{eq:action}
\small
a_t = \operatorname*{argmax}_{a \in \mathcal{A}} (p_{t,a})
\end{equation}

\begin{algorithm}[htbp]
	\caption{Persistent LinUCB Algorithm}
    \scriptsize
	\label{alg:linucb_mechanism}
	\LinesNumbered
	\SetAlgoLined
	\DontPrintSemicolon
	
	\SetKwInOut{Input}{Input}
	\SetKwFunction{FInit}{Initialize}
	\SetKwFunction{FSelect}{SelectAction}
	\SetKwFunction{FUpdate}{UpdateParameters}
	\SetKwProg{Fn}{Function}{:}{}
	
	\Input{Feature dimension $d=7$, Storage Path $\mathcal{P}$, Exploration $\alpha$}
	
	\BlankLine
	\tcp*[h]{Initialization \& Policy Loading}\label{line:initial-start}\\
	\Fn{\FInit{$\mathcal{P}$}}{
		\If{Model exists at path $\mathcal{P}$}{
			Load $\mathbf{A}, \mathbf{b}$ from $\mathcal{P}$\; \label{line:load_model}
		}
		\Else{
			\ForEach{$a \in \mathcal{A}$}{
				$\mathbf{A}_a \leftarrow \mathbf{I}_d$\;
				$\mathbf{b}_a \leftarrow \mathbf{0}_d$
			}
		}
	}\label{line:initial-end}
	
	\BlankLine
	\tcp*[h]{Policy Update \& Saving with any new observation \{$\mathbf{x}_t$, $a_t$, $r_t$\}}\\
	\Fn{\FUpdate{Context $\mathbf{x}_t$, Action $a_t$, Reward $r_t$}}{
		$\mathbf{A}_{a_t} \leftarrow \mathbf{A}_{a_t} + \mathbf{x}_t \mathbf{x}_t^\top$\; \label{line:update_A}
		$\mathbf{b}_{a_t} \leftarrow \mathbf{b}_{a_t} + r_t \cdot \mathbf{x}_t$\; \label{line:update_b}
		
		Save state $(\mathbf{A}, \mathbf{b})$ to path $\mathcal{P}$\;\label{line:save}
	}
	
	\BlankLine
	\tcp*[h]{Action Selection}\\
	\Fn{\FSelect{Context $\mathbf{x}_t$}}{
		\ForEach{$a \in \mathcal{A}$}{
			\If{$a = a_{\text{stop}}$}{
				$p_{t,a} \leftarrow 0.0$\;\label{line:stop}
			}
			\Else{
				$\hat{\boldsymbol{\theta}}_a \leftarrow \mathbf{A}_a^{-1} \mathbf{b}_a$\;\label{line:theta}
				$p_{t,a} \leftarrow \mathbf{x}_t^\top \hat{\boldsymbol{\theta}}_a + \alpha \sqrt{\mathbf{x}_t^\top \mathbf{A}_a^{-1} \mathbf{x}_t}$\; \label{line:ucb_calc}
			}
		}
		
		\Return $\operatorname*{argmax}_{a \in \mathcal{A}} (p_{t,a})$\; \label{line:argmax}
	}
\end{algorithm}

\subsection{Prompt Design and Tool Integration}\label{approach:mcp}
Following the LinUCB supervisor's selection of an action—either Default or Analysis—the framework dynamically generates a context-aware prompt to guide the coding agent's test generation process.

\subsubsection{Default Prompt}
If the supervisor selects the Default action ($a_{\text{def}}$), the framework injects the baseline prompt template shown in Fig. \ref{fig:prompt-template}. This template serves as the foundational system instruction, providing the coding agent with a strictly defined operational boundary and execution strategy. The prompt guides the coding agent to operate as an autonomous developer. It dynamically anchors the agent to the target class, establishes rigid success criteria, and supplies the necessary commands for environmental awareness. Crucially, rather than relying on a single attempt to generate tests, it enforces an iterative workflow, instructing the agent to actively identify gaps, synthesize tests, and independently repair failures based on execution feedback until code coverage is maximized.

\begingroup
\setlength{\abovecaptionskip}{2pt}
\providecommand{\beacon}{\textsc{Beacon}}        
\providecommand{\pfield}[1]{\textbf{#1}\enspace}  
\begin{figure}[htbp]
  \centering
  \begin{tcolorbox}[
    colback=gray!10!white, colframe=gray,
    boxrule=1.5pt, rounded corners,
    title=\textbf{Prompt Template},
    boxsep=3pt, top=4pt, bottom=4pt, left=5pt, right=5pt,
    fontupper=\footnotesize, fontlower=\footnotesize,
    before upper={\parindent=0pt \parskip=4pt\relax},
    before lower={\parindent=0pt \parskip=4pt\relax},
  ]

  \pfield{Task:} Automated unit test generation for a Java project. Increase branch coverage for the targeted class to the maximum while ensuring the project builds successfully and all tests pass.

  \pfield{Targeted Class:} \texttt{\{\{class\_name\}\}}

  \pfield{Workflow:} Identify existing test cases and coverage gaps, add new tests, execute and understand results, repair or remove failed tests, and verify that all tests pass.

  \pfield{Final Output:}\texttt{\{\{class\_name\}\}Test.java}

  \pfield{Success Criteria:} The project compiles without errors and all tests pass successfully.

  \pfield{Build \& Test Commands:}
  \begin{itemize}[nosep, left=0pt]
    \item Compile: \texttt{mvn clean compile}
    \item Test: \texttt{mvn test -Dtest=\{\{class\_name\}\}Test}
    \item JaCoCo coverage reports are under \texttt{target/jacoco}
  \end{itemize}

  \tcblower
  \pfield{(Analysis prompt) \beacon{}:} identifies coverage gaps, path constraints, and dependencies.
  \begin{itemize}[nosep, left=0pt]    
    \item \adjustbox{max width=\linewidth}{\texttt{get\_class\_analysis(project\_name, class\_name)}}   
  \end{itemize}

  \end{tcolorbox}
  \caption{Prompt template provided to a coding agent (condensed).}
  \label{fig:prompt-template}
\end{figure}
\endgroup

\subsubsection{Analysis Prompt and \beacon Integration}
When the LinUCB supervisor determines that a target class requires deeper structural reasoning, it selects the Analysis action ($a_{\text{ana}}$). In this state, our \toolname framework performs two critical operations to enable the Analysis action. First, it launches a program analysis tool \code{get\_class\_analysis} as an MCP server, making its program analysis capabilities seamlessly accessible to the coding agent. Second, it dynamically augments the Default prompt by appending the explicit MCP tool instructions (see the Analysis Prompt section in \autoref{fig:prompt-template}.
The complete Analysis prompt template is included in our replication package. Whereas the Default prompt relies entirely on the agent’s intrinsic reasoning over coverage reports, the Analysis prompt enhances its program analysis capabilities through integration with the \beacon tool.

For the implementation of \code{get\_class\_analysis}, we leverage the CFG 
as a middleware alongside coverage reports to extract uncovered execution paths. A significant challenge in this process is information overload: a highly complex class with low code coverage may contain hundreds of uncovered paths. Passing all of these paths simultaneously would overwhelm the context window of the underlying LLM and degrade the agent's focus. Conversely, providing only a single path per prompt would require excessive, high-latency LLM queries, wasting the model's reasoning capabilities. 

To balance this tradeoff, we design the tool to deliver paths in batches based on empirical observations. After extracting all paths containing uncovered lines and branches, \beacon sorts the methods by their number of uncovered statements. Within each method, the uncovered paths are subsequently sorted to identify the most critical path, defined as the path containing the highest number of uncovered statements, with any ties broken by selecting the shortest path. A single tool invocation provides the coding agent with a maximum of 10 under-covered methods, each accompanied by its most critical uncovered path. This targeted strategy allows the coding agent to maintain strict focus on generating tests for specific, high-priority coverage gaps.

Beyond execution paths, \beacon utilizes data flow analysis to resolve external method calls and their associated class types for each method under test. The coding agent can leverage this data to configure required mock objects or test setups, and to deduce the underlying behavior of the method by observing its external interactions. The extraction of uncovered paths and external dependencies directly addresses the testability bottlenecks quantified by the WMC and RFC metrics, which are integrated into the supervisor's contextual feature vector (Section \ref{approach:features}). An example of the resulting JSON payload is shown in~\autoref{fig:beacon_example}, illustrating the analysis of an under-covered \code{handleConcatenatedOptions} method from the \code{DefaultParser} class.

\begin{figure}
	\centering
    \scriptsize
	\begin{lstlisting}[language=json, basicstyle=\scriptsize\ttfamily, numbers=none, aboveskip=2pt, belowskip=2pt]
	{
		"current_coverage": {
			"line": 0.48,
			"branch": 0.32
		},
		"methods_under_test": [
			...
			{
				"method_declaration": {
					"name": "handleConcatenatedOptions",
					"signature": "void handleConcatenatedOptions (final String token)",
					"complexity": 4,
					"coverage_status": "no_coverage"
				},
				"path_with_uncovered_lines": [
					"for(int i = 1;i < token.length();i++) is True",
					"final String ch = String.valueOf(token.charAt(i));",
					"if(options.hasOption(ch)) is True",
					"handleOption(options.getOption(ch));",
					"if(currentOption != null && token.length() != i + 1) is True",
					"currentOption.addValueForProcessing(token.substring(i + 1));",
					"break;"
				],
				"external_calls": [
					"Option.addValueForProcessing(void)",
					"Options.getOption(String)",
					"Options.hasOption(String)"
				]
			},...]}
	\end{lstlisting}	
    \caption{Path Analysis by \beacon for \code{DefaultParser} class.}
	\label{fig:beacon_example}
\end{figure}

\subsection{Supervised Test Generation Workflow}\label{approach:workflow}
Algorithm \ref{alg:workflow} illustrates our workflow for a single target class, where the contextual bandit supervisor dictates the high-level strategy, while the agent executes the test generation.

The process begins by initializing the LinUCB policy ($\pi$), the core decision-making engine of the supervisor, by loading existing policy weights from the storage path ($\mathcal{P}$) via the \code{Initialize} function (Line \ref{line:init-policy}). Next, \toolname extracts the class metrics for the static portion of the context vector (Line \ref{line:init-metrics}). 

\begin{algorithm}[htbp]
    \caption{\toolname Workflow}
    \scriptsize
	\label{alg:workflow}
	\SetAlgoLined
	\DontPrintSemicolon
	
	\KwIn{Target Class ($\mathit{class\_name}$, $\mathit{project\_dir}$), Max Iterations $K_{max}$, Storage Path $\mathcal{P}$}
	\KwOut{Final Test Suite}
	\BlankLine
	\tcp*[h]{Load existing policy $\pi$, get (LOC, WMC, RFC)}\;
	$\pi \leftarrow \textbf{Algorithm~\ref{alg:linucb_mechanism}}.\text{Initialize}(\mathcal{P})$\;\label{line:init-policy}
	$metrics \leftarrow \text{GetClassMetrics}(class\_name, project\_dir)$\label{line:init-metrics}\;
	$k \leftarrow 0$\;
	
	\While{$k < K_{max}$}{\label{line:max-interation}
		$k \leftarrow k + 1$\;
		\BlankLine
		\tcp{Get coverage before generation}
		$Cov^L, Cov^B, Comp \leftarrow$ GetCoverage($class\_name$, $project\_dir$)\;\label{line:init-cov}
		\BlankLine
        \tcp{Store the valid test code before generation}
		$TestCode_{valid} \leftarrow \text{ExtractTestCode}(class\_name, project\_dir)$\;\label{line:init-testcode}
		
		\If{$Cov_L = 100\%$ \textbf{and} $Cov_B = 100\%$}{\label{line:saturated-start}
			\textbf{break} 
		}\label{line:saturated-end}
		
		$\mathbf{x}_k \leftarrow \text{ExtractFeatureVector}(metrics, Cov^L, Cov^B, Comp)$\;\label{line:context-vector}
		$a_k \leftarrow \pi.\text{SelectAction}(\mathbf{x}_k)$\;\label{line:select-action}
		
		\If{$a_k = a_\text{stop}$}{\label{line:stop-start}
			$\pi.\text{UpdateParameters}(\mathbf{x}_k, a_\text{stop}, 0)$\;
			\textbf{break}
		}\label{line:stop-end}
		
		$prompt, settings \leftarrow \text{PreparePromptAndSettings}(a_k, class\_name)$\;\label{line:prompt}
		$TestCode_{new}, Cost_k \leftarrow \text{ExecuteCodingAgent}(prompt, settings)$\;\label{line:agent-run}
		
		$Compiles, Passes \leftarrow$ EvaluateTests($class\_name$, $project\_dir$)\;\label{line:evaluate}
		\BlankLine
        \tcp{Get coverage after generation}
		$Cov'^L, Cov'^B, Comp'\leftarrow$ GetCoverage($class\_name$, $project\_dir$)\;
		$NoRegression \leftarrow (Cov'^L \ge Cov^L \textbf{ and } Cov'^B \ge Cov^B)$\;
		\eIf{$Compiles$ \textbf{and} $Passes$ \textbf{and} $NoRegression$}{
			$TestCode_{valid} \leftarrow TestCode_{new}$\label{line:accept}
		}{\label{line:rollback-start}
			$\text{RestoreTestFile}(TestCode_{valid})$ \tcp*[r]{Rollback}
			$(Cov'^L, Cov'^B, Comp') \leftarrow (Cov^L, Cov^B, Comp)$\;
		}\label{line:rollback-end}
		
		$r_k \leftarrow$ CalculateReward($Cov'^L$, $Cov'^B$, $Comp'$, $Cov^L$, $Cov^B$, $Comp$, $Cost_k$)\;\label{line:reward}
		$\pi.\text{UpdateParameters}(\mathbf{x}_k, a_k, r_k)$\;\label{line:adapt-policy}
	}
	\Return $TestCode_{valid}$\;
\end{algorithm}

We define the test generation process for a single target class as a \textit{generation trajectory}. To prevent infinite loops and optimize computational resources, the framework restricts each trajectory to a maximum number of iterations, $K_{max}$ (Line \ref{line:max-interation}). Note that while $k < K_{max}$ indexes the iterations within a single trajectory, the supervisor tracks its continuous adaptation using a global timestep, $t$, which accumulates all actions taken across all previous classes. At the start of each iteration $k$, \toolname evaluates the current line and branch coverage. Simultaneously, it extracts and stores the existing valid test code ($TestCode_{valid}$) to establish a safe baseline for potential rollbacks (Lines \ref{line:init-cov}--\ref{line:init-testcode}). If both line and branch coverage have already reached the threshold of $100\%$,
further generation is unnecessary, and the trajectory terminates early without requiring the supervisor's intervention (Lines \ref{line:saturated-start}--\ref{line:saturated-end}). 

If the trajectory continues, the framework dynamically constructs the context vector ($\mathbf{x}_k$) by fusing the static structural metrics with the current coverage status (Line \ref{line:context-vector}, detailed in Equation \ref{eq:context-vector}). The supervisor uses this context to select the current optimal action $a_k$ (Line \ref{line:select-action}). If the policy selects the $a_\text{stop}$ action, it signals that further generation efforts are likely futile; the process terminates, and the policy is updated with a reward of 0 (Lines \ref{line:stop-start}--\ref{line:stop-end}).

If the supervisor selects an active generation action (Default or Analysis), \toolname prepares the corresponding prompt and environment settings, instructing the agent to generate new tests (Lines \ref{line:prompt}--\ref{line:agent-run}). Upon completion, the framework records the token cost ($Cost_k$) and strictly evaluates the newly generated test suite. Because LLMs are prone to hallucinations that can introduce syntax errors or regress coverage, \toolname incorporates a guardrail mechanism. A new test suite is only accepted if it successfully compiles, passes all assertions, and does not regress existing coverage ($NoRegression$) (Lines \ref{line:evaluate}--\ref{line:accept}). If any of these conditions fail, the rollback mechanism discards the corrupted generation and restores the file to the previously saved $TestCode_{valid}$ state (Lines \ref{line:rollback-start}--\ref{line:rollback-end}). 

Finally, the framework calculates the reward signal ($r_k$) based on the net coverage improvement and the incurred token cost (Line \ref{line:reward}). This feedback is used to adapt the LinUCB policy weights for the selected action (Line \ref{line:adapt-policy}). This cycle repeats until the test suite is fully saturated, the supervisor selects the stop action, or the iteration limit $K_{max}$ is reached, at which point the final optimized test suite is returned.

\toolname operates in an online learning setting. Rather than learning from project-specific source code, the supervisor learns exclusively from contextual features derived from class-level testability metrics and generation feedback. After each generation action, the observed reward is used to update the LinUCB policy, thereby enabling the supervisor to continuously refine its decision-making strategy. To facilitate initial policy calibration, the supervisor is first exposed to a brief exploration period before transitioning to exploitation-guided decision making. During deployment, policy updates continue throughout test generation, allowing the supervisor to adapt to previously unseen classes. The specific training and evaluation used in our experiments are described in the next section.

\section{Evaluation}
Our study addresses the following research questions:

	
	
	

\begin{itemize}

    \item \textbf{RQ1:} How does \toolname's adaptive policy evolve during execution, and how does it respond to varying contextual characteristics?

    \item \textbf{RQ2:} What is the impact of \toolname's adaptive supervision on the effectiveness of coding agents?
    
    \item \textbf{RQ3:} How effective is \toolname compared to state-of-the-art non-agentic LLM-based approaches? 


\end{itemize}

\subsection{Experimental Setup}

\subsubsection{Dataset}
To evaluate \toolname, we selected the same Defects4J projects~\cite{just2014defects4j} used in the evaluation of the state-of-the-art LLM-based test generation approach, \panta~\cite{panta}. Unlike \panta, which relies solely on cyclomatic complexity, \toolname utilizes a set of class metrics (LOC, WMC, RFC). Since the primary objective of \toolname is to supervise coding agents on challenging test-generation tasks, we focus our evaluation on structurally complex classes. Following established thresholds~\cite{dcm-toolkit}, we retained classes with WMC $\geq 35$ or RFC $\geq 55$, and selected up to 10 qualifying classes per project to ensure project diversity while limiting the computational cost of repeated experimental runs. This process yielded a final dataset of 120 complex classes across 14 projects. The resulting dataset exhibits substantial structural complexity, with median values of 416.5 LOC, 68.0 WMC, and 80.5 RFC. Complete project-level metric distributions are provided in our replication package.

\subsubsection{Online Learning}\label{sec:learning-eval}

\begin{figure}
    \includegraphics[width=0.5\textwidth]{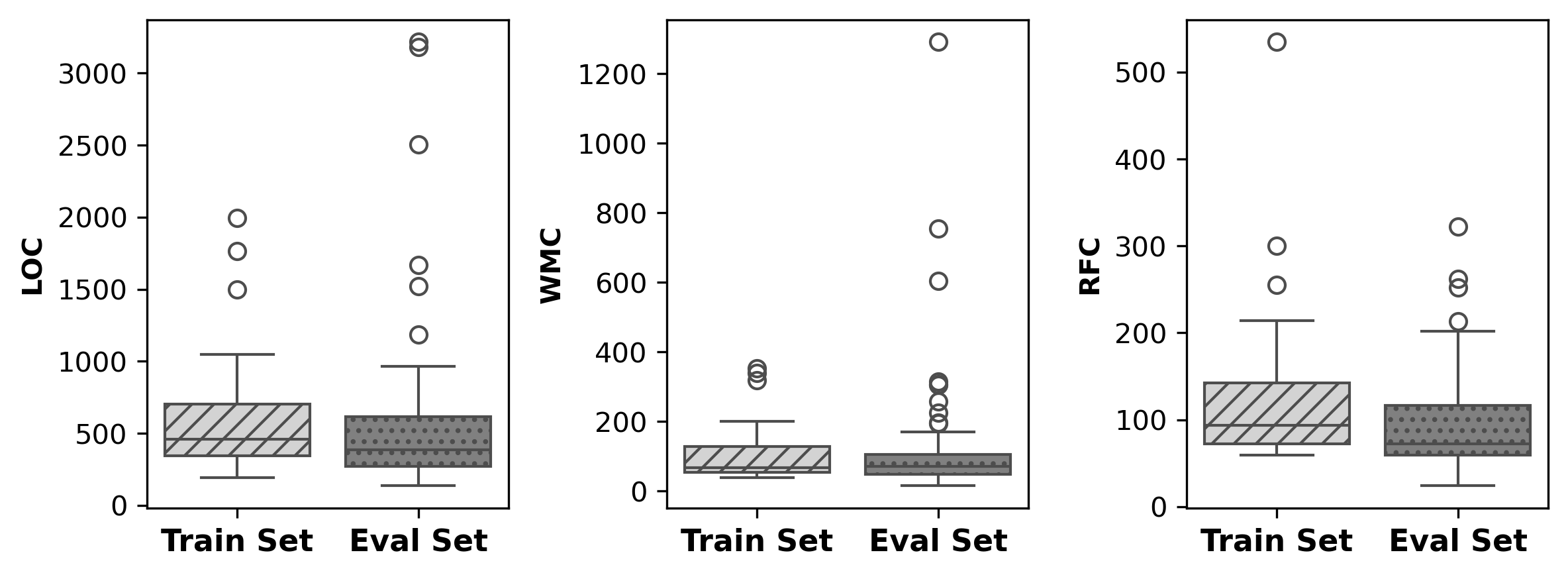}
    \captionof{figure}{Class metrics across the dataset partitions.}
    \label{fig:dataset-split}
\end{figure}

Because \toolname employs a contextual bandit design for continuous adaptation, we evaluate it using an online learning setup. Since the supervisor does not learn from project-specific source code, all 120 classes were randomly shuffled, regardless of project origin, and partitioned into a 40-class training set and an 80-class evaluation set. As shown in Figure~\ref{fig:dataset-split}, the two partitions exhibit comparable complexity distributions (median WMC of 67.5 and 70.0, respectively). Complete distribution statistics are available in our replication package.
The training phase consists of two stages. First, a \emph{random warm-up} phase uses the first 20 training classes for pure exploration. For each class, the supervisor selects generation actions uniformly at random for up to $K_{max}=10$ iterations, allowing unbiased collection of contextual and reward information. Second, an \emph{$\alpha$-decay} phase uses the remaining 20 training classes to transition from exploration to exploitation. During this stage, the supervisor selects actions using LinUCB while gradually decaying the exploration parameter from $\alpha=1.5$ to $\alpha=0.2$~\cite{linucb-analysis, linucb-guide, nguyen2024metallm}.
Following training, we evaluate \toolname on the remaining 80 unseen classes. During evaluation, the supervisor continues to update its policy online while maintaining a fixed exploration parameter of $\alpha=0.2$, simulating a deployment scenario in which supervision quality improves through continued interaction with new classes.

\subsubsection{Agents}

To address RQ1-–RQ3, we employ \geminicli~\cite{gemini-cli} (v0.35.3) as the primary coding agent within \toolname, configured with the \code{gemini-3.1-flash-lite} model (temperature = 1.0). We selected this model due to its support for tool calling, relatively low inference cost~\cite{gemini-model}, and stable availability~\cite{gemini-429-error} throughout the experimental period. For RQ2, we also evaluate \toolname with \claudecode~\cite{claude-code} (v2.1.173) using \code{claude-4-5-haiku}, a model selected for its comparable speed and cost profile. 

\toolname supports both augmenting existing test suites and generating tests from scratch. To ensure a fair comparison across all research questions, we evaluate the latter scenario, where no existing test cases are provided and coverage for each target class starts at zero. Our experiments were conducted between April--June 2026.

\subsection{RQ1: Execution Dynamics}

To address RQ1, we investigate the execution dynamics of \toolname's policy to understand what the supervisor learns and how its behavior evolves. 

\subsubsection{Learning Progress and Policy Adaptation}

To evaluate the learning efficacy, we analyzed the lifetime average cumulative reward $\bar{R}_t = \frac{1}{t} \sum_{i=1}^{t} r_i$ across all timesteps $t \in T$. This step-wise formulation provides a high-fidelity representation of the policy's learning progress over time ~\cite{poole2010artificial}.

\begin{figure}
	\includegraphics[width=0.5\textwidth]{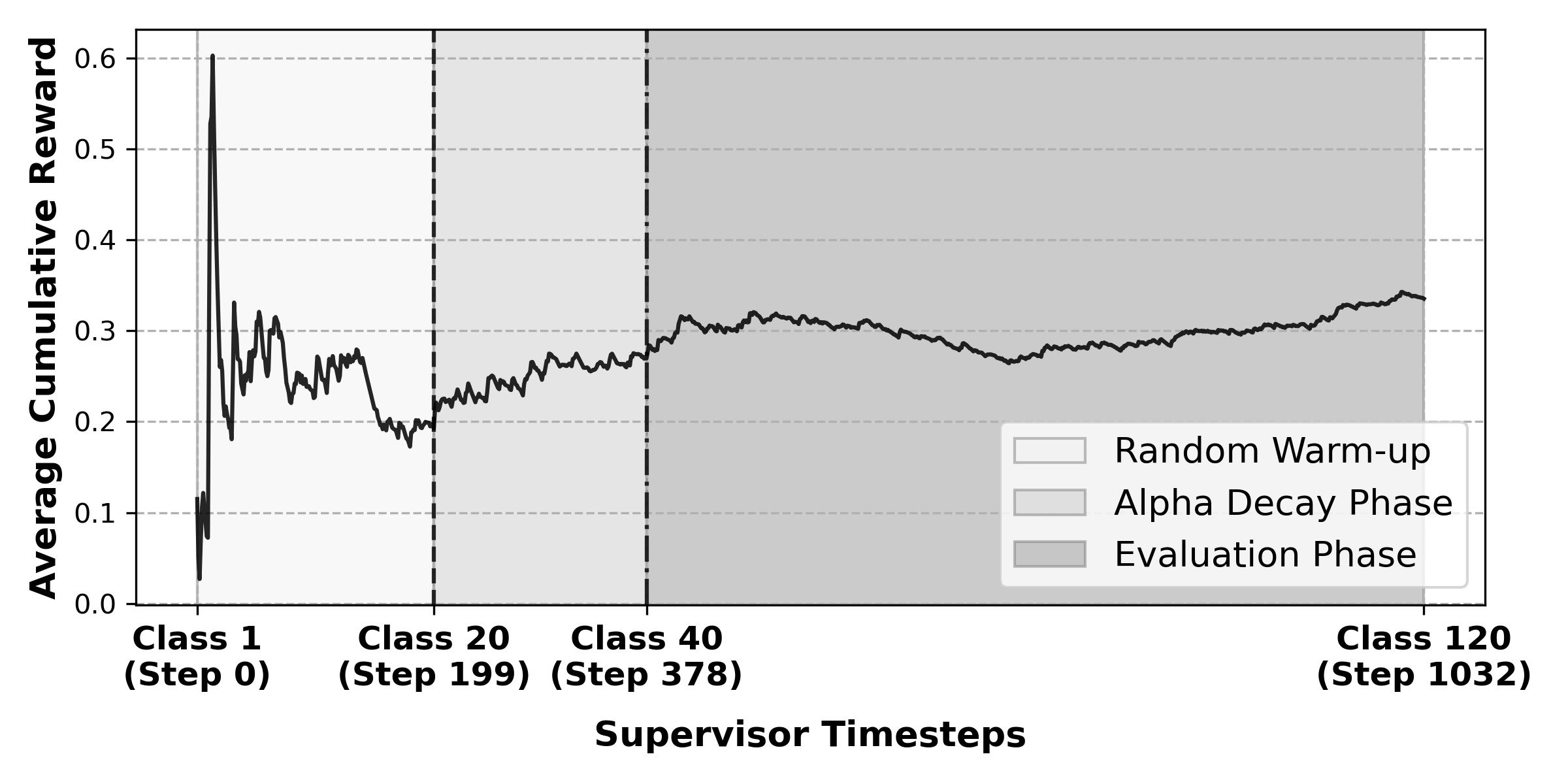}
	\caption{Lifetime Average Cumulative Reward}
	\label{learning-curve}
\end{figure}

As shown in Figure~\ref{learning-curve}, the average cumulative reward remains near 0.2 during the random warm-up phase (Classes 1–20 | Timesteps 0–199), where actions are selected primarily for exploration. During the subsequent $\alpha$-decay phase (Classes 21–40 | Timesteps 200–378), the supervisor progressively shifts toward exploitation, resulting in a steady increase in average cumulative reward to 0.3. This improvement suggests that the supervisor has learned to make more effective routing decisions based on the observed testing context.

Following training, the exploration rate stabilizes at $\alpha = 0.2$, prioritizing exploitation while maintaining online adaptation. During evaluation (Classes 41–120), the average cumulative reward remains stable despite a brief decline on several highly complex classes and subsequently resumes its upward trend. This result suggests that the supervisor continues to improve its decision-making over time.

\subsubsection{Intra-Trajectory Analysis of Supervisor Behavior}

\begin{figure}
	\includegraphics[width=0.5\textwidth]{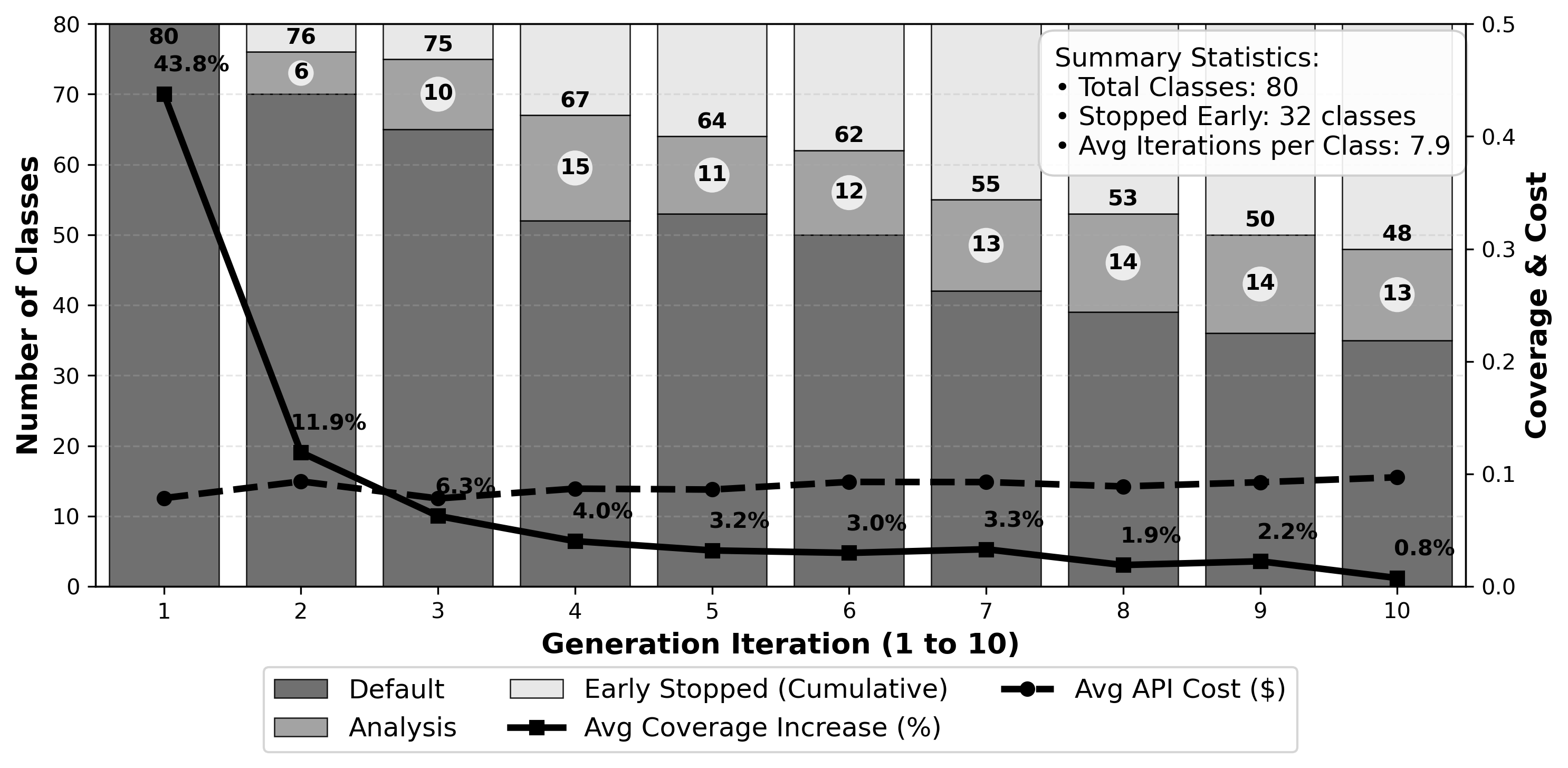}
	\caption{Execution Dynamics of \toolname Generation Trajectories}
	\label{fig:intra-trajectory}
\end{figure}

Following the assessment of overall learning progress, we analyze the behavior of the learned policy on the 80 evaluation classes. 
Figure~\ref{fig:intra-trajectory} details the execution dynamics within trajectories across the 10-iteration generation horizon. 

The largest coverage gains occur early in the trajectory. Iteration 1 yields an average coverage increase of 43.8\%, followed by an additional 11.9\% in Iteration 2. During this phase, the supervisor relies almost exclusively on the \textit{Default} action, with only six classes invoking \textit{Analysis} in Iteration 2. This behavior suggests that when generation begins from zero coverage, the standard \textit{Default} prompt is generally sufficient to rapidly capture easily reachable coverage.

As the trajectory progresses, the rate of coverage improvement naturally decreases as the remaining uncovered paths become increasingly complex. Despite that, the supervisor continues to achieve positive coverage gains throughout the entire 10-iteration horizon.
It consistently allocates a subset of classes to \textit{Analysis}, with 11–15 classes invoking deeper program analysis per iteration between Iterations 4--10. This behavior suggests that the learned policy reserves \textit{Analysis} for classes where additional coverage remains achievable despite increasing structural complexity and diminishing returns from standard generation.

As returns diminish, the supervisor increasingly issues \textit{Stop} actions when neither \textit{Default} nor \textit{Analysis} yields sufficient reward. The cumulative number of early-terminated classes steadily increases, ultimately reaching 32 of the 80 evaluation classes (40\%) by Iteration 10. This aligns with the reward formulation described in Section~\ref{approach:reward}, which balances coverage gains against computational cost. Because the API cost of \code{gemini-3.1-flash-lite} remains low (below \$0.10 USD per iteration), the cost component contributes only a modest penalty. Consequently, the learned policy prioritizes extracting additional coverage when meaningful gains remain available while avoiding premature termination.

By Iteration 10, the average coverage gain drops to 0.8\%, while 40\% of classes have already been terminated early. Together, these observations suggest that the predefined $K_{max}=10$ cutoff effectively bounds generation cost without sacrificing substantial additional coverage. Overall, the average trajectory length is 7.9 iterations. By dynamically orchestrating different actions throughout the generation process, \toolname demonstrates its ability to adapt action selection to the contextual characteristics of the target class, balancing coverage improvement against computational cost.

\subsubsection{Impact of Contextual Features on Action Selection}
To further understand the supervisor's decision-making mechanism, we analyze how the contextual state of the target class influences the \toolname supervisor’s action selection. In our formulation, context comprises both static class metrics and dynamic code coverage. To structure this analysis, we categorize the generation trajectories by the types of actions selected  and the total length of the trajectory before termination ($\le 5$ vs.\ $> 5$). Specifically, \textit{Default only} refers to generation trajectories where the supervisor relies exclusively on the \textit{Default} action across all iterations. Conversely, \textit{with Analysis} denotes trajectories where the supervisor invokes \textit{Analysis} action in at least one iteration before termination. Table~\ref{tab:bandit_context} presents the average values of these contextual features across the classes in each category.

As shown in the upper half of Table~\ref{tab:bandit_context}, the static complexity of a target class heavily influences the supervisor's decision to invoke the \textit{Analysis} action. Trajectories categorized as \textit{Default only} are associated with smaller, simpler classes (average LOC of 469.3, WMC of 98.5, and RFC of 90.3) that yield high average coverage ($Cov^L = 85.9\%$, $Cov^B = 78.7\%$). For these classes, the supervisor correctly learns that the \textit{Default} prompt is sufficient to achieve adequate code coverage. By contrast, the supervisor learns to deploy the \textit{Analysis} action for significantly larger and more complex classes (average LOC of 755.1, WMC of 163.9, and RFC of 104.8). Due to this inherent complexity, the final achieved code coverage for these classes is lower (76.6\% line and 66.7\% branch coverage). This indicates that the supervisor correctly determines when the \textit{Analysis} action is required, i.e.,  when a target class possesses complex logic, and the standard \textit{Default} mode is no longer sufficient to increase coverage.

\begin{table}
    \centering
    \scriptsize
    \caption{Impact of Contextual State on Action Selection (Avg).}
    \label{tab:bandit_context}
    \begin{tabular}{@{} l l c c c c c @{}}
        \toprule
        \multicolumn{2}{@{}l}{\textbf{Trajectory Category}} & \textbf{LOC} & \textbf{WMC} & \textbf{RFC} & $\boldsymbol{Cov^L}$ & $\boldsymbol{Cov^B}$ \\
        \midrule
        \textit{Active Action} & Default Only  & 469.3 & 98.5  & 90.3  & \textbf{85.9\%} & \textbf{78.7\%} \\
                               & With Analysis & \textbf{755.1} & \textbf{163.9} & \textbf{104.8} & 76.6\% & 66.7\% \\
        \midrule
        \textit{Iterations}    & $\le 5$       & 443.0 & 83.2  & 87.3  & \textbf{90.9\%} & \textbf{85.8\%} \\
                               & $> 5$         & \textbf{601.4} & \textbf{131.4} & \textbf{97.5}  & 80.4\% & 71.4\% \\
        \bottomrule
    \end{tabular}
\end{table}

The bottom half of Table~\ref{tab:bandit_context} reveals how contextual state dictates the trajectory length and the timing of the \textit{Stop} action. Short trajectories ($\le 5$ iterations) are strongly associated with simpler classes (average LOC of 443.0, WMC of 83.2, and RFC of 87.3), where high code coverage (90.0\% line and 85.8\% branch coverage) is easily achieved in fewer iterations. Once coverage saturates on these simpler classes, the supervisor executes an early \textit{Stop} to prevent wasteful iterations that yield no further reward. In contrast, longer trajectories ($> 5$ iterations) correlate with higher static complexity (average LOC of 601.4, WMC of 131.4, and RFC of 97.5) and lower overall coverage levels (80.4\% line and 71.4\% branch coverage). For these challenging classes, the supervisor persists longer, maintaining active generation and analysis to incrementally extract remaining coverage gains until diminishing returns force termination.

\subsection{RQ2: Supervised vs. Unsupervised Agents}

We compare against four standalone-agent baselines that operate without the supervisor. The first two baselines represent single-shot executions: \textit{Agent (Default)}, which utilizes the standard \textit{Default} prompt, and \textit{Agent (Analysis)}, which utilizes the \textit{Analysis} prompt alongside the \beacon tool.
Finally, to ensure a rigorously fair comparison against \toolname's multi-iteration trajectories, with an average of 7.9 iterations across the evaluation set, we introduce two iterative baselines: \textit{Agent (Default, 8 Iters)} and \textit{Agent (Analysis, 8 Iters)}. These configurations force the agent to generate tests for eight consecutive iterations using their respective setups. The baselines and \toolname use \geminicli as the underlying coding agent. All prompts are included in our replication package.

Table~\ref{tab:baseline_comparison} reports the per-class average values across the evaluation set for line coverage ($Cov^L$), branch coverage ($Cov^B$), mutation score (mutation \%), cost, generation iterations, and overall execution time. As shown, the single-shot baselines demonstrate that deep analysis is not strictly advantageous for initial test generation. \textit{Agent (Default)} achieves comparable or slightly superior coverage ($Cov^L$ of 50.5\%, $Cov^B$ of 43.8\%) compared to \textit{Agent (Analysis)} ($Cov^L$ of 49.6\%, $Cov^B$ of 43.9\%), while operating at a lower cost (\$0.079 vs. \$0.085). This validates \toolname's learned behavior observed in RQ1, where the supervisor universally selected the \textit{Default} action for the first iteration to efficiently extract easily achievable coverage. Furthermore, these results highlight the inherent limitations of single-shot generation: restricted to a single iteration, both baselines fail to achieve high overall code coverage, falling nearly 31\% behind \toolname in branch coverage.

\begin{table}
    \centering
    \scriptsize
    \caption{Comparison of (un)supervised agents (\geminicli).}
    \label{tab:baseline_comparison}
    \setlength{\tabcolsep}{1.5pt}
    \begin{tabular}{@{} l c c c c c c @{}}
        \toprule
        \textbf{Configuration} & $\boldsymbol{Cov^L}$ & $\boldsymbol{Cov^B}$ & \textbf{Mutation \%} & \textbf{Cost} & \textbf{Iterations} & \textbf{Time (M)}\\
        \midrule
        
        Agent (Default)  & 50.5\% & 43.8\% & - &\$0.079 & 1.0 & 1.5\\
        Agent (Analysis) & 49.6\% & 43.9\% & - &\$0.085 & 1.0 & 2.1\\
        \midrule
        
        Agent (Default, 8 Iters)  & 77.3\% & 70.2\% & 59.6\% &\$0.797 & 8.0 & 16.5\\
        Agent (Analysis, 8 Iters) & 74.7\% & 67.5\% & 58.1\% &\$0.567 & 8.0 & 14.5\\
        \textbf{\toolname} & \textbf{82.8\%} & \textbf{74.7\%} & \textbf{64.7\%} &\textbf{\$0.693} & \textbf{7.9} & \textbf{15.1}\\
        
        \bottomrule
    \end{tabular}
\end{table}

The \textit{Agent (Analysis, 8 Iters)} baseline achieves a line coverage of 74.7\% and a branch coverage of 67.5\%. We use a one-sided Wilcoxon signed-rank test ($\alpha = 0.05$) to evaluate differences between \toolname and the baseline, as it is a robust nonparametric method well-suited to paired, non-normally distributed software metrics. This test confirms that \toolname yields statistically significant improvements, achieving 82.8\% line coverage (+8.1\%, $p < 0.001$) and 74.7\% branch coverage (+7.2\%, $p < 0.001$) within an equivalent average iteration footprint. Similarly, when compared with the \textit{Agent (Default, 8 Iters)} baseline, \toolname achieves 5.5\% higher line coverage ($p = 0.007$) and 4.5\% higher branch coverage ($p = 0.004$). 

In addition to line and branch coverage, we measure mutation score using Pitest (v1.17.0). \toolname achieves an average mutation score of 64.7\%,  significantly outperforming both the \textit{Agent (Analysis, 8 Iters)} (58.1\%, $p<0.001$) and \textit{Agent (Default, 8 Iters)} (59.6\%, $p=0.002$) baselines. In terms of both cost and runtime, \toolname falls between the two baselines. It averages \$0.693 and 15.1 minutes, which is slightly higher than \textit{Agent (Analysis, 8 Iters)} (\$0.567, 14.5 min) but lower than \textit{Agent (Default, 8 Iters)} (\$0.797, 16.5 min). These results indicate that \toolname's significant gains in coverage and mutation scores come without substantial computational overhead. Ultimately, the comparison with both iterative baselines demonstrates that the supervisor's ability to dynamically interleave actions based on real-time structural context outperforms a static, homogeneous agent loop.

Additionally, to assess the impact of the underlying coding agent, we compare \toolname when paired with \claudecode and \geminicli. Unsupervised, \claudecode achieves substantially higher initial coverage (84.3\% line and 79.0\% branch) than \geminicli (50.5\% and 43.8\%, respectively), but at a considerably higher Iteration 1 cost (\$0.518 vs. \$0.079). Consequently, \toolname adapts its supervision policy by terminating \claudecode early for 95.0\% of classes, resulting in short trajectories averaging 2.1 iterations and only 10.0\% of classes invoking \textit{Analysis}. In contrast, \geminicli's lower cost and greater opportunity for coverage improvement lead to longer trajectories (7.9 iterations on average) and more frequent \textit{Analysis} usage (33.8\%). Ultimately, \toolname paired with \claudecode achieves higher final coverage (90.3\% line and 84.9\% branch) alongside a 75.0\% mutation score at a total cost of \$1.143, while with \geminicli it reaches competitive coverage (82.8\% line and 74.7\% branch) and a 64.7\% mutation score for \$0.693. These results demonstrate that \toolname dynamically adapts its supervision strategy to the capabilities and cost profile of the underlying coding agent.

\subsection{RQ3: Non-Agentic State-of-the-Art Comparison}

We compare \toolname against two state-of-the-art LLM-based test generation approaches, \panta~\cite{panta} and \symprompt~\cite{symprompt}, which combine program analysis with prompt-based strategies. While \panta is publicly available~\cite{panta-repo}, \symprompt is not; therefore, we use the high-fidelity \symprompt replication provided by the \panta authors~\cite{panta-repo}. To ensure a fair comparison, all approaches are evaluated on the same 80-class dataset using the identical underlying model (\code{gemini-3.1-flash-lite}, temperature = 1.0). This isolates the impact of \toolname’s supervised agentic workflow relative to static prompting-based generation pipelines.


\begin{figure}
	\includegraphics[width=0.5\textwidth]{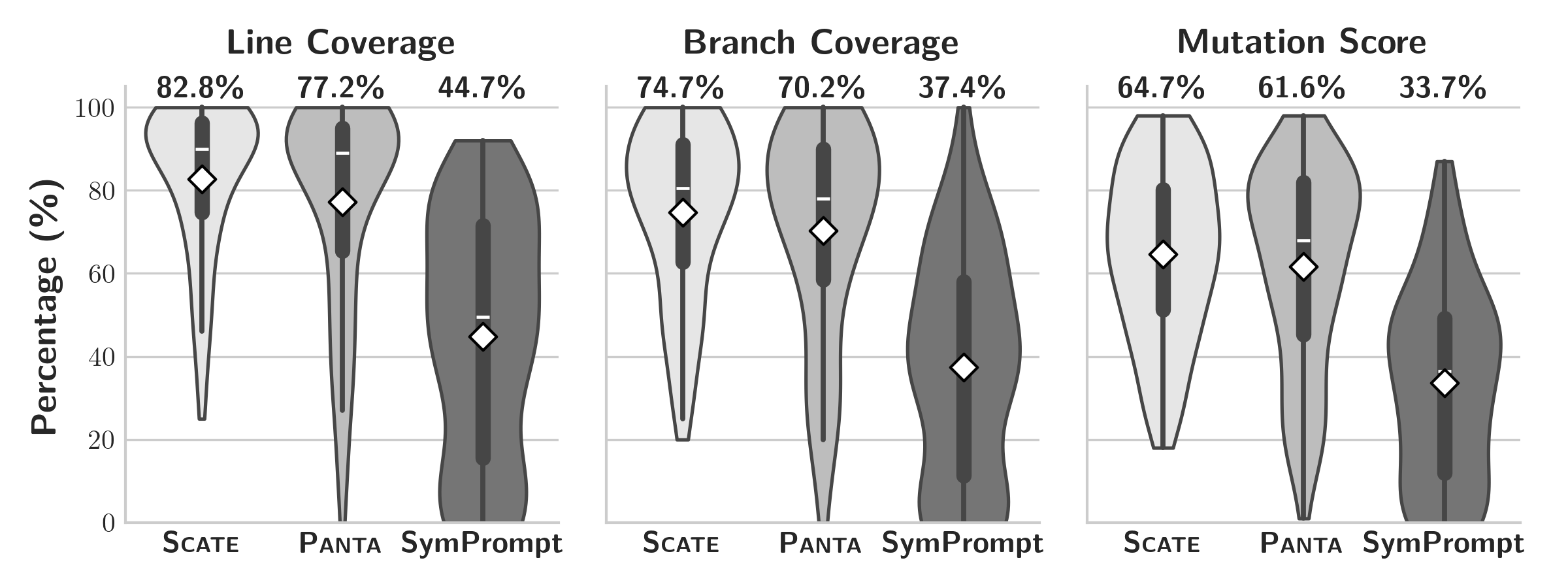}
	\caption{Comparing \toolname with \panta and \symprompt}
	\label{panta-comparison}
\end{figure}

Figure~\ref{panta-comparison} shows the distributions of coverage and mutation score using violin plots with embedded boxplots. Statistical significance is assessed using a one-sided Wilcoxon signed-rank test ($\alpha = 0.05$). As illustrated, \toolname consistently outperforms both \panta and \symprompt across all three metrics. For Line Coverage, \toolname achieves a mean of 82.8\%, surpassing \panta (77.2\%, $p=0.007$) and \symprompt (44.7\%, $p<0.001$). Beyond these higher averages, \toolname's top-heavy distribution indicates a high concentration of classes achieving near-perfect coverage. Furthermore, \toolname establishes a substantially higher empirical minimum compared to both baselines. In contrast, \symprompt exhibits a severely bottom-heavy distribution extending down to 0\%, reflecting a high frequency of low-coverage outcomes. Moving to Branch Coverage, the data reflects a similar structural advantage: \toolname maintains a superior mean of 74.7\% and a concentrated upper quartile, whereas \panta (70.2\%, $p=0.009$) and \symprompt (37.4\%, $p<0.001$) exhibit pronounced lower tails, demonstrating that static prompting frequently fails to navigate intricate path logic, while \toolname successfully resolves these complex conditions to achieve high branch coverage. 

Most critically, the Mutation Score validates \toolname's effectiveness beyond mere structural execution by evaluating the suite's fault-detection capability. \toolname achieves a leading mean of 64.7\% and maintains a robust empirical floor consistently near or above the 20\% threshold. Conversely, both \panta (61.6\%, $p=0.017$) and \symprompt (33.7\%, $p<0.001$) exhibit a high density of classes falling below 20\%. Ultimately, these results demonstrate \toolname's superior capability in generating high-quality test suites that maximize both code coverage and overall test effectiveness.

\section{Discussion}
\header{Application} 
Because the supervisor relies entirely on source-code-independent numerical metrics rather than specific syntax, the learned policy generalizes well beyond the Defects4J dataset. To support out-of-the-box usage, our replication package includes pre-trained policies for \geminicli and \claudecode. Users can also train \toolname for other coding agents (e.g., \qwencode) or different underlying LLMs.

\header{Efficiency and Cost} Training a customized \toolname policy is a one-time investment. For our 40-class training set, policy learning required 13 hours and \$48.40 using \geminicli, compared with 22.7 hours and \$127.00 using \claudecode. Once trained, generating a test suite required an average of 15.1 minutes and \$0.69 per class with \geminicli, and 11.1 minutes and \$1.14 with \claudecode. Generation time scaled with class complexity (1–49 minutes for \geminicli and 2–60 minutes for \claudecode). Because \toolname integrates directly with standard coding agents, users with commercial subscriptions can typically absorb both the training and generation cost through their existing usage quotas.

\header{Agentic vs. Non-Agentic} Agentic and non-agentic test generation represent different design philosophies. Non-agentic approaches, such as \panta and \symprompt, operate on narrow contexts using carefully engineered prompting pipelines, resulting in lower inference cost but limited reasoning capability. In contrast, agentic approaches leverage repository-wide context, iterative planning, tool invocation, and multi-step reasoning, enabling them to address complex dependencies and substantially improve test quality at the expense of additional computation. Our results demonstrate that this tradeoff yields significantly higher code coverage and mutation scores despite the increased inference cost.

\header{Threats To Validity} A threat to internal validity is the stochasticity of both the coding agent and the contextual bandit. We mitigate this by normalizing rewards and contextual features, randomly shuffling the training and evaluation classes, evaluating only after empirical policy convergence, and reporting aggregate results with statistical significance testing. Together, these measures reduce the influence of stochastic variation and class-ordering bias.

External validity may be limited by our use of a single benchmark (Defects4J) and programming language (Java). However, both the supervisor formulation and the underlying coding agents are language-agnostic. Because Defects4J projects are publicly available, LLM pretraining contamination is also a potential concern. To ensure fair comparisons, all evaluated approaches use the same underlying models and configurations, and we emphasize relative improvements over baselines rather than absolute performance. Finally, although training and evaluation classes originate from the same projects, data leakage is mitigated because the supervisor learns exclusively from project-independent numerical features rather than source code.

\header{Reproducibility} Code and datasets will be made publicly available upon publication.
\section{Related Work}
Automated unit test generation has been a longstanding research topic, with approaches historically falling into two main categories: search-based ~\cite{evosuite:fse11, fraser2013evosuite, pynguin, lukasczyk2020automated} and learning-based ~\cite{atlas, tufano2020unit, tufano-pretrained-22}. The rapid evolution of Large Language Models (LLMs) has led to their widespread adoption for tasks related to automated test generation ~\cite{alshahwan2024automated, kang2023large, cedar:icse23, lemieux:codamosa:icse23, utgen, chen2024chatunitest, pan2025aster, fu2026fusing}. Empirical studies have also been conducted to evaluate the test generation capabilities of different LLMs in various programming languages ~\cite{schafer2023empirical, yang2024evaluation, siddiq2024using, wang2024testeval, yuan2023no, elhaji:test-generation-using-copilot:thesis23, shang:empirical-fine-tuning-unit-testing:issta25}. To tackle complex code, static tools such as HITS~\cite{wang2024hits}, JUnitGenie~\cite{liao:path-sensitive-unit-test:ase25}, and \symprompt~\cite{symprompt} rely on program analysis, prompting LLMs with context extracted as method slices or execution paths, whereas CoverUp~\cite{altmayer2025coverup} overcomes their inability to augment existing test suites by leveraging dynamic runtime coverage. \panta~\cite{panta} combines dynamic coverage with static control-flow analysis to generate new tests and augment existing ones. While other techniques primarily focus on improving code coverage, MutGen~\cite{wang:mutation-guided-test-generation:tse26} shifts its objective toward maximizing the mutation score.


Contextual bandits~\cite{li2010contextual} are a specialized form of reinforcement learning designed to map contexts to actions that maximize an immediate reward. Within the broader AI community, this approach has recently guided dynamic LLM routing in tools such as MetaLLM~\cite{nguyen2024metallm}, LLMBandit~\cite{li2025llm}, and MixLLM~\cite{wang2025mixllm}. Concurrently, general AI frameworks such as SupervisorAgent~\cite{lin2025stop} have introduced real-time supervision to improve efficiency in multi-agent systems, though they do not utilize a contextual bandit formulation. To the best of our knowledge, we are the first to introduce the notion of adaptive supervision in the software engineering domain, proposing a contextual bandit-driven supervisor dedicated to maximizing the efficiency of automated test generation.
\section{Conclusion}

We introduced \toolname, a supervisor-orchestrated agentic framework for automated unit test generation. \toolname addresses the \emph{lazy generation} phenomenon in modern coding agents, where agents prematurely terminate instead of generating tests for complex program logic. By formulating supervision as a contextual bandit problem, \toolname replaces human intervention with an adaptive, metric-driven policy that continuously evaluates runtime coverage and structural testability to decide whether to continue standard generation, invoke program analysis, or terminate the generation process. Through this adaptive supervision, \toolname balances code coverage and computational cost. Our evaluation on a diverse dataset constructed from Defects4J demonstrates that \toolname significantly outperforms both standalone coding agents and state-of-the-art non-agentic approaches, improving line and branch coverage by 32.3\% and 30.9\%, respectively, over the standalone agent baseline. These results demonstrate that adaptive supervision is a key ingredient for realizing the full potential of agentic software testing.
	

\bibliographystyle{IEEEtran}
\interlinepenalty=10000
\bibliography{main}
	
\end{document}